# On the efficiency of computational imaging with structured illumination

T.E. Gureyev[1,2,3,4],*, D.M. Paganin[2], A. Kozlov[1], Ya.I. Nesterets[5,3] and H.M. Quiney[1]

[1] ARC Centre of Excellence in Advanced Molecular Imaging, The University of Melbourne, Parkville, VIC 3010, Australia
[2] School of Physics and Astronomy, Monash University, Clayton, VIC 3800, Australia
[3] School of Science and Technology, University of New England, Armidale, NSW 2351, Australia
[4] Data61, Commonwealth Scientific and Industrial Research Organisation, Clayton, VIC 3168, Australia
[5] Manufacturing, Commonwealth Scientific and Industrial Research Organisation, Clayton, VIC 3168, Australia
* Corresponding author: timur.gureyev@unimelb.edu.au

## Abstract

A generic computational imaging setup is considered which assumes sequential illumination of a semi-transparent object by an arbitrary set of structured illumination patterns. For each incident illumination pattern, all transmitted light is collected by a photon-counting bucket (single-pixel) detector. The transmission coefficients measured in this way are then used to reconstruct the spatial distribution of the object's projected transmission. It is demonstrated that the squared spatial resolution of such a setup is usually equal to the ratio of the image area to the number of linearly independent illumination patterns. If the noise in the measured transmission coefficients is dominated by photon shot noise, then the ratio of the spatially-averaged squared mean signal to the spatially-averaged noise variance in the "flat" distribution reconstructed in the absence of the object, is equal to the average number of registered photons when the illumination patterns are orthogonal. The signal-to-noise ratio in a reconstructed transmission distribution is always lower in the case of non-orthogonal illumination patterns due to spatial correlations in the measured data. Examples of imaging methods relevant to the presented analysis include conventional imaging with a pixelated detector, computational ghost imaging, compressive sensing, super-resolution imaging and computed tomography.

## 1. Introduction

The duality between decomposition and synthesis is one of the most pervasive and fruitful ideas in mathematical physics. Key to rendering precise the decomposition-synthesis duality is the concept of expansions utilising a superposition of basis objects drawn from a complete set. The process of decomposition involves breaking down an object belonging to a suitably wide class of possible objects, expressing it as a weighted superposition of objects in the complete basis set. The inverse process of synthesis takes the weighting coefficients referred to in the previous sentence, using them to synthesise the desired object from the said coefficients.

Geometrically, the decomposition-synthesis duality sets up an evident correspondence with linear algebra, a correspondence that is developed further in the formalism of functional analysis [1]. This is particularly useful when the underpinning equations, such as the Maxwell equations for electromagnetic waves [2] or the parabolic equation of paraxial wave optics [3], are linear differential equations. The process of decomposition may thereby be viewed as determining all possible projections of a vector in a specified function space (with the set of all possible vectors in the function space being associated with the set of all possible objects to be decomposed), with the process of synthesis corresponding to constructing a given vector (representing a particular object) from the knowledge of each of its projections [1].

A wide class of optical imaging scenarios may be viewed from this perspective. An obvious example is pixelated CCD cameras, in which an arbitrary N × N pixel image may be synthesised as a superposition of one-pixel "basis" images, each of which has only one of the pixels uniformly illuminated [4]. Other examples relevant to optical imaging include the plane-wave (Fourier basis) [5], other complete-eigenfunction expansions such as Hermite-Gauss and Gauss-Laguerre bases [3], Floquet expansions [6], the Huygens construction as embodied in e.g. the Rayleigh-Sommerfeld diffraction integrals [7] and the convolution formulation of Fresnel diffraction theory [8], the Green's-function formalism for linear shift-variant (LSI) imaging systems and the convolution formalism for such systems [5], computational ghost imaging [9,10], ghost imaging using random speckle bases [11,12], the wavelet decomposition of optical images [13], the singular-value decomposition of computed tomography [14] etc. The basis elements in this far from exhaustive list range from being maximally localised (as in the pixel basis, together with its limit case given by the Dirac-delta basis) to being maximally delocalised (as in the plane-wave basis, many modal decomposition bases, and in convolution-type propagator formalisms embodying the Huygens construction). The wavelet basis corresponds to basis elements with a level of localisation that is intermediate between the pixel basis or the Dirac-delta basis, and the other bases mentioned above.

The choice of basis elements, in a complete set utilised in the context of solving a given optical problem, is arbitrary. A corollary of this arbitrariness is the power it imparts to choose a basis in which the solution to a given optical problem assumes a particularly convenient, transparent, tractable or otherwise desirable form. Thus, for example, a plane-wave basis may be chosen in a free-space diffraction context on account of the analytical power of the associated machinery of Fourier analysis together with the extreme numerical efficiency of the fast Fourier transform algorithm [5], a spatially random speckle field basis may be chosen in a ghost imaging context on account of its amenability to compressive sensing concepts [15], a wavelet basis may be chosen when it is useful to have a sparse representation of an optical image, a modal basis may be useful in an optical communication context where only a small number of such modes is likely to be excited [16], and so forth. For example, in the case of parallel-beam computed tomography, if the incident illumination patterns are structured radially in the shape of suitable Zernike polynomials, then a simple and fast reconstruction of the object can be provided in terms of Chebyshev polynomials and the measured transmission coefficients [14].

A second corollary of this arbitrariness, of the choice of complete basis in the setting of the optical decomposition-synthesis duality, is the opportunity it affords to study optical imaging from a perspective that is not limited to a particular choice of complete basis. This is the perspective adopted by the present paper, in addressing the particular questions of spatial resolution and signal-to-noise ratio for intensity-linear optical imaging using arbitrary complete bases. While most of the results of the present study are generally applicable to arbitrary bases as applied to optical imaging, we also draw some conclusions specific to the comparison between the localised pixel basis associated with direct imaging, and less localised bases associated with indirect forms of imaging such as computational ghost imaging [9].

There is an evident analogy between inline holography, viewed as the two-step process of image recording followed by reconstruction [17], and the generic means of indirect intensity imaging considered here. While this latter problem, which is the core topic of the present paper, also considers imaging as a two-step process of recording followed by reconstruction, in our case the object to be reconstructed is the projected transmission distribution of a semi-transparent object which can be obtained by means of a linear transformation of the intensity registered by the detector, rather than both the intensity and phase of a complex disturbance.

We close this introduction with an outline of the remainder of the paper. Section 2 gives a formalism for describing a generic intensity-linear imaging setup whose ultimate purpose is to determine a spatially resolved estimate for an object transmission distribution obtained from a set of integral transmitted intensity measurements by a single-pixel detector when the object is illuminated by a sequence of different structured illumination patterns. Special cases that can be described by this model include but are not limited to direct imaging using a pixelated detector, and indirect imaging of the impinging intensity distribution using computational imaging [9,10], ghost imaging [11,12], coded aperture imaging [18], etc. The present formalism is developed utilising ghost-imaging terminology, particularly in utilising the term "bucket signal" for what is in essence the total scattering cross-section [11,12], but as emphasised earlier the domain of applicability of the formalism is much broader than ghost imaging. This section considers a means for determining the point-spread function (PSF) associated with a given set of illumination patterns, both for the general case of a position-dependent PSF (Green's function), and the special case (where applicable) of a position-independent PSF. Section 3 then considers the associated question of spatial resolution for our rather general imaging setup, special cases of which include all of the previously-mentioned forms of both direct and indirect imaging. Signal-to-noise-ratio (SNR) in the reconstructed object transmission distribution is then considered in Section 4. The obtained expression for the SNR is factorised into a product of a function depending solely upon the photon statistics of the illuminating field, and a function depending solely upon the object and basis functions. The analysis is then presented from the perspective of a recently-introduced "intrinsic quality characteristic" [19,20]. This characteristic is invariant with respect to a broad class of intensity-linear shift-invariant transformations of imaging systems [20], and it effectively quantifies the optical quality of an imaging system (i.e. its efficiency of utilization of photons for imaging of the illuminated object). Sections 5-7 draw specific comparison between the fully localised and non-localised bases, from the perspectives of SNR and spatial resolution of the reconstructed object transmission distribution, and the previously-mentioned imaging quality metric. We conclude with Section 8, where we discuss a classification of imaging systems on the basis of transformation of SNR and spatial resolution between the measurement (image) and the reconstruction (object) spaces.

## 2. A computational imaging setup

Consider Fig.1, which shows an imaging setup where two-dimensional light intensity patterns $I_m(\mathbf{r})$, $\mathbf{r} = (x, y)$, $m = 1,2,\dots,M$, are used to illuminate, in sequence, a thin semi-transparent object which is described by a deterministic dimensionless real non-negative transmission function $X(\mathbf{r})$. The illuminating patterns $I_m(\mathbf{r})$ are assumed to have been implicitly integrated over some fixed exposure time, and therefore they are expressed in the units of energy density (energy per unit area, J/m$^2$). During each exposure, a "bucket" (single-pixel) photon-counting detector, with efficiency $\eta$ (expressed in photons per Joule) and a sufficiently large sensitive surface region $\Omega$, collects all the transmitted light. The signal measured by the detector during the exposure with a given illumination pattern $I_m(\mathbf{r})$ is a single non-negative dimensionless number $a_m = \eta \iint\limits_{\Omega} X(\mathbf{r}) I_m(\mathbf{r}) d\mathbf{r}$, which corresponds to the number of registered photons. Such bucket signals may be measured at any distance from the object as long as the attenuation in the gap between the object and the detector is negligible. The goal of the imaging experiment is to reconstruct (or at least, to approximate) the unknown object transmission function $X(\mathbf{r})$ from a set of measured "bucket" coefficients $a_m$, $m = 1,2,\dots,M$.

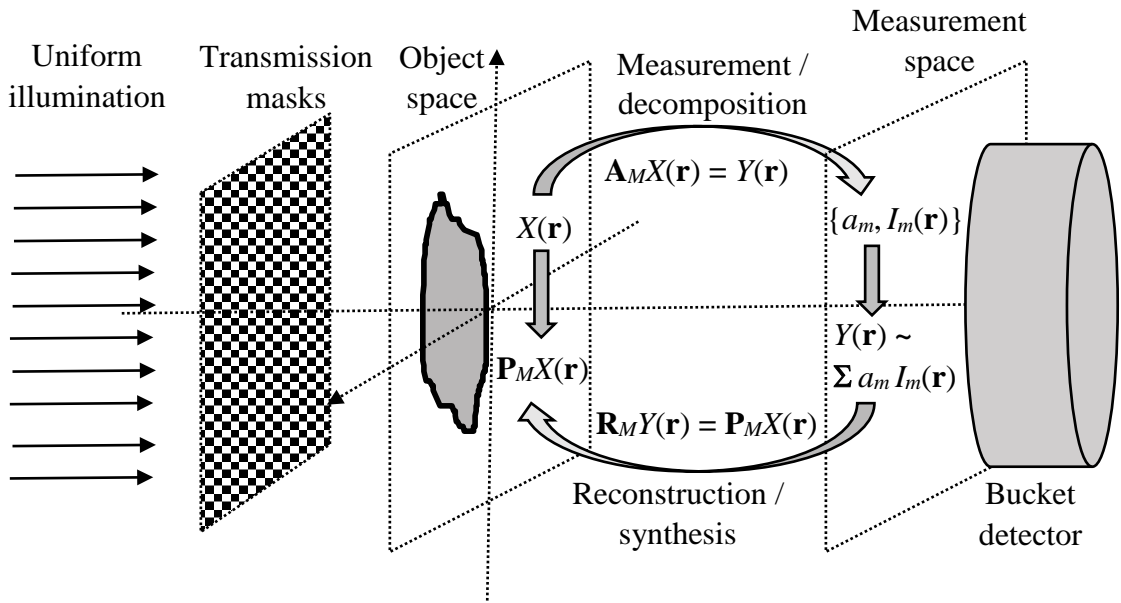

Fig.1. Diagram of the imaging setup considered in the present paper.

The described imaging scheme (see Fig.1) may correspond e.g. to computational ghost imaging [9-12], possibly in combination with compressive sensing [15,21], or other imaging methods with structured illumination and a single-pixel detector. In the present paper, we are primarily interested in questions about the spatial resolution and signal-to-noise ratio that can be achieved in these types of imaging setup. Note, however, that our context is much broader than that of ghost or computational imaging, notwithstanding our use of the term "bucket" for the photon detector.

In order to maximize the class of unknown transmission functions $X(\mathbf{r})$ that could be accurately reconstructed from the measurement of coefficients $a_m$, $m = 1,2,\ldots,M$, the illumination patterns $I_m(\mathbf{r})$ should be linearly independent, i.e. none of the functions $I_m(\mathbf{r})$ should be representable as a linear combination of other functions $I_{m'}(\mathbf{r}), m' \neq m$, with constant coefficients. If the set of patterns $I_m(\mathbf{r})$ is not linearly independent, the linear subspace spanned by the functions $I_m(\mathbf{r})$, $m = 1,2,\ldots,M$, will have a dimension $M' < M$, and the class of reconstructable transmission functions will be narrower compared to the linearly independent case. In practice, in some methods relevant to our study, the set of illumination patterns can be linearly dependent, but that only represents a straightforward additional technicality for the model that we consider. For simplicity, in the present paper we assume that the set $I_m(\mathbf{r})$, $m = 1,2,\ldots,M$, is linearly independent.

In order to calculate the effect of photon shot noise on the coefficients $a_m$ measured by the bucket detector, we assume that all illumination patterns $I_m(\mathbf{r})$ are obtained using the same incident illumination which is sufficiently monochromatic and has uniform intensity distribution, $I_{in}(\mathbf{r}) = I_{in}$, transmitted through different spatially varying transmission masks $T_m(\mathbf{r})$, i.e. $I_m(\mathbf{r}) = I_{in}T_m(\mathbf{r})$ with $0 \leq T_m(\mathbf{r}) \leq 1$. The mask transmission distributions $T_m(\mathbf{r})$ are assumed to be deterministic (static) and dimensionless. A typical case is represented by the functions $T_m(\mathbf{r}) = \exp[-\mu L_m(\mathbf{r})]$, where $\mu$ is the linear attenuation coefficient of the mask material and $L_m(\mathbf{r})$ is the spatial distribution of the mask's projected thickness along the direction of the incident illumination. For simplicity we assume that the integrals $t^2 \equiv \dfrac{1}{|\Omega|} \iint\limits_{\Omega} T_m^2(\mathbf{r}) d\mathbf{r}$ are the same for all $m = 1,2,\ldots,M$, where $|\Omega|$ denotes the area of $\Omega$. The illumination patterns $I_m(\mathbf{r})$ are also assumed to be deterministic in our model, while the bucket coefficients $a_m$ are independent random variables with Poisson statistics, as discussed in detail below.

We will also consider an optional condition that the set of illumination patterns $I_m(\mathbf{r})$ may satisfy.

<u>Condition 1</u>. The constant function, such that $f(\mathbf{r}) = 1$ everywhere in $\Omega$, belongs to the linear space spanned by the functions $I_m(\mathbf{r})$, $m = 1, 2, ..., M$, i.e. $1 = \sum_{m=1}^{M} \alpha_m I_m(\mathbf{r})$ for some constant coefficients $\alpha_m$.

If Condition 1 does not hold for a given set of illumination patterns $I_m(\mathbf{r})$, $m = 1, 2, ..., M$, it means that the constant functions are linearly independent with respect to the set $I_m(\mathbf{r})$. In this case, it should usually be easy to add an extra illumination pattern $I_{M+1}(\mathbf{r}) = const$ (e.g. just a flat-field illumination with no mask, $I_{M+1}(\mathbf{r}) = 1$) to the set in order to satisfy Condition 1.

We will see below that Condition 1 expresses a form of an energy conservation law. In practical terms, Condition 1 means that any uniform transmission function can be reconstructed from the measured bucket coefficients $a_m$, which is equivalent to the statement that the coefficients $c_m$ in Condition 1 can be expressed as linear combinations of the bucket coefficients $\eta \iint_{\Omega} I_m(\mathbf{r}) d\mathbf{r}$. The ability to reconstruct uniform transmission functions represents a desirable property expected from a well-designed imaging system. Condition 1 is usually satisfied in most real systems. This natural condition will be shown to lead to some convenient simplifications in our analysis of the spatial resolution and SNR below.

Mathematically, we will consider only square-integrable illumination patterns and transmission functions and view them as vectors from the linear space $L_2(\Omega)$ of square-integrable functions over $\Omega$. The scalar product of two vectors $\boldsymbol{V} = V(\mathbf{r})$ and $\boldsymbol{U} = U(\mathbf{r})$ is defined as

$$< \boldsymbol{V}, \boldsymbol{U} > \equiv \frac{1}{|\Omega|} \iint_{\Omega} V(\mathbf{r}) U(\mathbf{r}) d\mathbf{r} . \qquad (1)$$

We will denote the corresponding vector length as $\| \boldsymbol{V} \| = (< \boldsymbol{V}, \boldsymbol{V} >)^{1/2}$. It will be convenient for us to work with dimensionless "illumination vectors" $\boldsymbol{W}_m \equiv \eta |\Omega| I_m(\mathbf{r})$, $m = 1, 2, ..., M$. Note that the bucket coefficients can be represented as $a_m = < \boldsymbol{X}, \boldsymbol{W}_m >$, where $\boldsymbol{X} \equiv X(\mathbf{r})$.

If the set of vectors $\boldsymbol{W}_m$, $m = 1, 2, ..., M$, is linearly independent, they can always be orthonormalized [22,23]. Such orthonormalization is not unique. One can use e.g. the polar decomposition for this purpose. Let $\mathbf{W}$ be the matrix with elements $(W_{nm}) = < \boldsymbol{E}_n, \boldsymbol{W}_m >$ in some orthonormal basis $\{E_n\}$, i.e. the matrix that consists of vectors $W_m = WE_m$ as its columns. The polar decomposition of $W$ can be written as $W = V(W^\dagger W)^{1/2}$, where the superscript "$\dagger$" denotes transposition, $(W^\dagger W)^{1/2}$ is a positive-definite matrix, and $V$ is an orthogonal matrix. In particular, all columns of the matrix $V = W(W^\dagger W)^{-1/2}$ are orthonormal with respect to the scalar product Eq.(1). Therefore, the set of vectors $V_m = VE_m$ represents an orthonormalization of vectors $W_m$, $m = 1, 2, \ldots, M$:

$$V_m = \sum_{m'=1}^{M} q_{mm'} W_{m'}, \text{ and } <V_m, V_{m'}> = \delta_{mm'}, m, m' = 1, 2, \ldots, M, \tag{2}$$

where $\delta_{mm'}$ is the Kronecker delta. The coefficients $q_{mm'}$ in Eq.(2) correspond to the matrix $Q = (W^\dagger W)^{-1/2}$. The matrix $Q^2 = (W^\dagger W)^{-1}$ with elements $q_{mm'}^{(2)} = \sum_{l=1}^{M} q_{ml} q_{lm'}$ can be used to construct a basis $\{U_m\}$ that is biorthogonal to $\{W_m\}$:

$$U_m = \sum_{m'=1}^{M} q_{m'm} V_{m'} = \sum_{m'=1}^{M} q_{mm'}^{(2)} W_{m'}, \ <U_m, W_{m'}> = \delta_{mm'}, m, m' = 1, 2, \ldots, M. \tag{3}$$

The bases $\{V_m\}$ and $\{U_m\}$ are used below for reconstructing the object transmission function from the measured bucket coefficients.

Let us supplement the set of vectors $V_m$, $m = 1, 2, \ldots, M$, with suitable additional vectors to form a complete orthonormal basis $\{V_m\}$, $m = 1, 2, \ldots, \infty$, in the Hilbert space $L_2(\Omega)$ with the scalar product defined by Eq.(1). Obviously, the whole functional space $L_2(\Omega)$ can be represented as a sum of the sub-space $V_M(\Omega)$, spanned by the first $M$ basis vectors, and its orthogonal complement: $L_2(\Omega) = V_M(\Omega) \oplus V_M^\perp(\Omega)$.

The measured "signal" $\{a_m, W_m\}$ can be represented as a vector $Y = \sum_{m=1}^{M} a_m W_m$ in the subspace $V_M(\Omega)$. When the vectors $W_m$, $m = 1, 2, \ldots, M$, are linearly independent, the coefficients $a_m$ in this representation of the signal vector are unique, i.e. if $Y = \sum_{m=1}^{M} a_m W_m = \sum_{m=1}^{M} a'_m W_m$, then $a_m = a'_m$ for all $m$. It is useful to introduce the (linear) measurement operator $A_M$ which maps the object transmission function $X(\mathbf{r})$ into the measured signal vector:

$\mathbf{A}_M X = Y = \sum_{m=1}^{M} a_m W_m$. By definition, the reconstruction (synthesis) operator $\mathbf{R}_M$ is equal to

the inverse of the restriction of the measurement operator $\mathbf{A}_M$ to the subspace $V_M(\Omega)$. Let us show that $\mathbf{R}_M Y = \sum_{m=1}^{M} a_m U_m$, and that such reconstruction represents a projection of the object transmission function $X = X(\mathbf{r})$ onto the subspace $V_M(\Omega)$.

Consider a projection operator from $L_2(\Omega)$ onto the sub-space $V_M(\Omega)$. This operator acts on an arbitrary vector $X$ from $L_2(\Omega)$ according to the expression $\mathbf{P}_M X = \sum_{m=1}^{M} b_m V_m$,

$b_m \equiv <X, V_m>$. Using Eq.(3), it is then easy to verify that

$$\mathbf{P}_M X = \sum_{m=1}^{M} b_m V_m = \sum_{m=1}^{M} c_m W_m = \sum_{m=1}^{M} a_m U_m , \quad b_m = \sum_{m'=1}^{M} q_{mm'} a_{m'} , \quad c_m = \sum_{m'=1}^{M} q_{mm'}^{(2)} a_{m'} . \qquad (4)$$

In particular, Eq.(4) shows that $\mathbf{R}_M Y = \mathbf{P}_M X$, i.e. this reconstruction indeed corresponds to the projection of the object transmission function onto the vector subspace spanned by the illumination patterns. Note that, if the original illumination vectors $W_m$, $m = 1, 2, ..., M$, are themselves orthogonal and all have the same length, i.e. $<W_m, W_{m'}> = w^2 \delta_{mm'}$ with some constant $w > 0$, then we can take $U_m = w^{-1} V_m = w^{-2} W_m$, $m = 1, 2, ..., M$, $q_{mn} = w^{-1} \delta_{mn}$ and $c_m = w^{-1} b_m = w^{-2} a_m$, greatly simplifying Eq.(4).

Let us see under which conditions the reconstructed function $\mathbf{R}_M Y(\mathbf{r}) = \mathbf{P}_M X(\mathbf{r})$ represents a good approximation to the object transmission function $X(\mathbf{r})$. It is easy to verify that the operator $\mathbf{P}_M$ is a projector, i.e. $\mathbf{P}_M^2 = \mathbf{P}_M$ [24], and it can be represented as a linear integral operator

$$\mathbf{P}_M X(\mathbf{r}) = \frac{1}{|\Omega|} \iint_{\Omega} G_M(\mathbf{r}, \mathbf{r}') X(\mathbf{r}') d\mathbf{r}' \qquad (5)$$

with the kernel (Green's function)

$$G_M(\mathbf{r}, \mathbf{r}') = \sum_{m=1}^{M} U_m(\mathbf{r}) W_m(\mathbf{r}') = \sum_{m=1}^{M} V_m(\mathbf{r}) V_m(\mathbf{r}').$$ (6)

From a physical perspective, the operator $\mathbf{P}_M$ projects an arbitrary object $X$ to its approximation as a linear combination of the $M$ independent basis vectors. This generalised form of filtering, which is consistent with the concept of reconstructing an object to within a finite resolution, may naturally be considered in geometric terms as a projection since it discards any components of the object which are orthogonal to all of the basis vectors. This intuitively aligns with the formal definition of an operator $\mathbf{A}$ being a projector if $\mathbf{A}^2 = \mathbf{A}$, since applying a projection more than once has the same effect as applying it once.

In some imaging methods, e.g. in computational ghost imaging [9,10], the illumination patterns are often chosen in such a way, that the following optional condition is satisfied.

*Condition 2*. The set of illumination patterns is such that the corresponding Green's function is shift-invariant (spatially stationary), at least approximately, i.e.
$G_M(\mathbf{r}+\mathbf{h}, \mathbf{r}'+\mathbf{h}) = G_M(\mathbf{r}, \mathbf{r}')$ for any vector $\mathbf{h}$, provided that $\mathbf{r}$, $\mathbf{r}'$, $\mathbf{r}+\mathbf{h}$ and $\mathbf{r}'+\mathbf{h}$ lie within $\Omega$.

If Condition 2 holds, the Green's function can be represented as a function of the difference between the two coordinates: $G_M(\mathbf{r}, \mathbf{r}') = P_M(\mathbf{r}-\mathbf{r}')$, where the corresponding function $P_M(\mathbf{r}) = \sum_{m=1}^{M} V_m(0) V_m(\mathbf{r})$ acts as a PSF [25]:

$$\mathbf{P}_M X(\mathbf{r}) = \frac{1}{|\Omega|} \iint_\Omega P_M(\mathbf{r}-\mathbf{r}') X(\mathbf{r}') d\mathbf{r}'.$$ (7)

Note that Condition 1 implies that the spatial average of the PSF is equal to 1:
$1 = \mathbf{P}_M 1(\mathbf{r}) = \frac{1}{|\Omega|} \iint_\Omega P_M(\mathbf{r}-\mathbf{r}') d\mathbf{r}'$, and hence the convolution with the PSF does not change the total transmission through the object
$\iint_\Omega \mathbf{P}_M X(\mathbf{r}) d\mathbf{r} = \iint_\Omega \frac{1}{|\Omega|} \iint_\Omega P_M(\mathbf{r}-\mathbf{r}') X(\mathbf{r}') d\mathbf{r}' d\mathbf{r} = \iint_\Omega X(\mathbf{r}') d\mathbf{r}'$. In this sense, Condition 1 implies a natural conservation law which states that the projection onto sub-space $V_M(\Omega)$ preserves the total transmission value of the object.

The measurement operator $\mathbf{A}_M$ can also be represented as a linear integral operator:

$$\mathbf{A}_M X(\mathbf{r}) = \frac{1}{|\Omega|} \iint_\Omega A_M(\mathbf{r},\mathbf{r}') X(\mathbf{r}') d\mathbf{r}' \qquad (8)$$

with the kernel (Green's function)

$$A_M(\mathbf{r},\mathbf{r}') = \sum_{m=1}^M W_m(\mathbf{r}) W_m(\mathbf{r}') . \qquad (9)$$

Indeed, if we substitute Eq.(9) into Eq.(8), and take $X(\mathbf{r}) = \sum_{m=1}^M a_m U_m(\mathbf{r})$, we obtain:

$$\frac{1}{|\Omega|} \iint_\Omega \sum_{m=1}^M W_m(\mathbf{r}) W_m(\mathbf{r}') X(\mathbf{r}') d\mathbf{r}' = \sum_{m=1}^M W_m(\mathbf{r}) < \sum_{m'=1}^M a_m U_{m'}, W_m > = \sum_{m=1}^M a_m W_m(\mathbf{r}), \text{ i.e.}$$

$\mathbf{A}_M X(\mathbf{r}) = Y(\mathbf{r})$, as required. Similarly, the reconstruction operator $\mathbf{R}_M$ can be represented as a linear integral operator with the kernel $R_M(\mathbf{r},\mathbf{r}') = \sum_{m=1}^M U_m(\mathbf{r}) U_m(\mathbf{r}')$. When the illumination patterns are orthogonal, such that $< W_m, W_{m'} > = w^2 \delta_{mm'}$, both the measurement and the reconstruction operator are proportional to the identity operator in $V_M(\Omega)$:

$\mathbf{A}_M \mathbf{P}_M X(\mathbf{r}) = w^2 \mathbf{P}_M X(\mathbf{r})$ and $\mathbf{R}_M Y(\mathbf{r}) = w^{-2} Y(\mathbf{r})$.

## 3. Spatial resolution

In view of Eq.(7), in order for the reconstructed function $\mathbf{R}_M Y(\mathbf{r}) = \mathbf{P}_M X(\mathbf{r})$ to be a good approximation for $X(\mathbf{r})$, for an arbitrary imaged object $X(\mathbf{r})$, the PSF $P_M(\mathbf{r})$ should approximate Dirac's delta function. Equation (6) then becomes a completeness (closure) relation: $|\Omega|^{-1} \sum_{m=1}^M V_m(\mathbf{r}) V_m(\mathbf{r}') \cong \delta(\mathbf{r}-\mathbf{r}')$ [24]. In general, however, the projection operator $\mathbf{P}_M$ effectively blurs, i.e. degrades the spatial resolution of, any function $X(\mathbf{r})$ on which it acts. We will see below that when the basis vectors correspond to the Fourier harmonics, the blurring due to the convolution with the relevant PSF corresponds to truncation of the Fourier decomposition of the object transmission function to the linear combination of the first $M$ Fourier components (low-pass filtration).

One can see that in general the degree of blurring in Eq.(5), i.e. the width of the reconstructed response to a delta-function like input signal $X(\mathbf{r}')$, can be different for different points $\mathbf{r}$. However, when the Green's function is shift-invariant, the width of the response to a localised input is the same at every point in $\Omega$, being equal to the width of the PSF. This width corresponds to the spatial resolution of the imaging setup.

According to [20,26,27], a convenient measure of the width of a function $g(\mathbf{r})$ can be given by the expression:

$$(\Delta_2 r)[g] \equiv \frac{\iint g(\mathbf{r})d\mathbf{r}}{\left(\iint g^2(\mathbf{r})d\mathbf{r}\right)^{1/2}} .\tag{10}$$

Unlike the more conventional measure of width, based on the variance of a function, the definition in Eq.(10) works well with the formalism of vector decomposition over an orthogonal basis, as demonstrated below. At the same time, the width defined according to Eq.(10) produces values which are fully consistent with the natural understanding of the width of a function in the case of Gaussians, Lorentzians, rectangular (uniform) and other popular distributions [20]. It is also important for the following to note that the definition of spatial resolution in Eq.(10) is "optimistic", in the sense that for any integrable function $g(\mathbf{r})$ the following inequality holds: $(\Delta_2 r)[g] \le (3\pi^{1/2} / 2)(\Delta r)[g]$, where

$\left((\Delta r)[g]\right)^2 \equiv \iint |\mathbf{r} - \bar{\mathbf{r}}|^2 | g(\mathbf{r})| d\mathbf{r} / \iint | g(\mathbf{r})| d\mathbf{r}$ [20]. In other words, the spatial resolution estimated in accordance with Eq.(10) is always finer than or equal to the more conventional spatial resolution, defined via the spatial variance of the PSF, multiplied by the constant $3\pi^{1/2} / 2 \cong 2.67$. Details about the relationship between the two definitions of the spatial resolution (width of the PSF) can be found in [20].

Firstly we consider the spatial resolution of the function $g_{\mathbf{r}'}(\mathbf{r}) = G(\mathbf{r}', \mathbf{r}) = G(\mathbf{r}, \mathbf{r}')$ which depends on the argument $\mathbf{r}'$ as a parameter. We will assume for simplicity that Condition 1 is satisfied. The corresponding results without Condition 1 can be derived in the same way, but the expressions are more complex. When the constant function 1 lies in the space $V_M(\Omega)$, we have $\mathbf{P}_M 1 = 1$, and hence $\iint_\Omega g_{\mathbf{r}'}(\mathbf{r})d\mathbf{r} = \iint_\Omega G_M(\mathbf{r}', \mathbf{r})d\mathbf{r} = |\Omega| (\mathbf{P}_M 1)(\mathbf{r}') = |\Omega|$. Next, using the orthonormality of vectors $V_m$, we obtain:

$\iint_\Omega g_{\mathbf{r}'}^2(\mathbf{r})d\mathbf{r} = \iint_\Omega G_M^2(\mathbf{r}', \mathbf{r})d\mathbf{r} = \sum_{m=1}^{M} V_m^2(\mathbf{r}')\iint_\Omega V_m^2(\mathbf{r})d\mathbf{r} = |\Omega| \sum_{m=1}^{M} V_m^2(\mathbf{r}')$. Therefore,

$$(\Delta_2 r)[g_\mathbf{r}] = \left( |\Omega| / \sum_{m=1}^{M} V_m^2(\mathbf{r}) \right)^{1/2} .\tag{11}$$

Obviously, this resolution may in general be different at different points $\mathbf{r}$ in $\Omega$. Note that the spatial average of the function $f_M(\mathbf{r}) = \sum_{m=1}^{M} V_m^2(\mathbf{r})$ over $\Omega$ is equal to $M$, due to the normalization of vectors $V_m$: $\frac{1}{|\Omega|} \iint_\Omega \sum_{m=1}^{M} V_m^2(\mathbf{r}) d\mathbf{r} = \sum_{m=1}^{M} \|V_m\|^2 = M$. If the Green's function is shift-invariant, the function $f_M(\mathbf{r})$ is constant in $\Omega$, because

$f_M(\mathbf{r}) = \sum_{m=1}^{M} V_m^2(\mathbf{r}) = G_M(\mathbf{r},\mathbf{r}) = G_M(0,0) = f_M(0)$, and hence $f_M(\mathbf{r}) = M$ for any $\mathbf{r}$. Therefore, in the shift-invariant case, $\iint_\Omega g_\mathbf{r}^2(\mathbf{r}) d\mathbf{r} = M |\Omega|$ and $(\Delta_2 r)[g_\mathbf{r}] = (|\Omega|/M)^{1/2}$. As mentioned previously, in the shift-invariant case, the Green's function can be represented as a PSF, $G_M(\mathbf{r},\mathbf{r}') = P_M(\mathbf{r}-\mathbf{r}')$, and so the width of the Green's function in this case is equal everywhere to the width of the PSF, which determines the uniform spatial resolution:

$$(\Delta_2 r)[P_M] = (|\Omega|/M)^{1/2}. \qquad (12)$$

The square of the width of the PSF can be interpreted as the "effective pixel" area. Equation (10) shows that the square of the spatial resolution of the considered computational imaging system using $M$ illumination patterns $I_m(\mathbf{r})$, $m = 1,2,\ldots,M$, which are linearly independent and satisfy Conditions 1 and 2, is always equal to the image area divided by the number of illumination patterns. This simple result is consistent, for example, with the expected spatial resolution of a 2D imaging system with a low-pass filter retaining only the first $M$ Fourier harmonics. It is interesting that this intuitive result holds for the much more general class of bases considered here, than just e.g. the Fourier basis. We consider some relevant illustrative examples in more detail below.

By calculating the average width of the Green's function $A_M(\mathbf{r},\mathbf{r}') = \sum_{m=1}^{M} W_m(\mathbf{r}) W_m(\mathbf{r}')$ using Eq.(10) as above, it is possible to verify that the spatial resolution of the measured signal $Y(\mathbf{r}) = \sum_{m=1}^{M} a_m W_m(\mathbf{r})$ is equal to

$(\Delta_2 r)[A_M] = \left( |\Omega| \sum_m \sum_{m'} <W_m><W_{m'}><W_m,W_{m'}> / \sum_m \sum_{m'} <W_m,W_{m'}>^2 \right)^{1/2}$. In the orthogonal case, i.e. when $<W_m,W_{m'}> = w^2 \delta_{mm'}$, we have $<1> = <\sum_{m=1}^{M} <W_m> U_m(\mathbf{r})> = w^{-2} \sum_{m=1}^{M} <W_m>^2$ as a consequence of Condition 1. Using this fact it easy to verify that $(\Delta_2 r)[A_M]$ becomes equal to $(|\Omega|/M)^{1/2}$. Comparing this result with Eq.(12), we see that in the case of

orthogonal illumination patterns the average spatial resolution is the same in the measurement and in the object spaces.

## 4. Signal-to-noise ratio

Now let us consider the signal-to-noise ratio (SNR) of the reconstructed distribution described by Eq.(4) or, equivalently, Eq.(5). Here the "signal" at each point $\mathbf{r}$ is formally defined as a mean (ensemble averaged) value, $\overline{\mathbf{P}_M X}(\mathbf{r}) = \sum_{m=1}^{M} \overline{a}_m U_m(\mathbf{r})$, of the function $\mathbf{R}_M Y(\mathbf{r}) = \mathbf{P}_M X(\mathbf{r})$ reconstructed multiple times using Eq.(4) with coefficients $a_m$ measured under identical experimental conditions; $\overline{a}_m$ are the mean values of the measured coefficients. The random character of the measurements of coefficients $a_m$ is assumed to be determined by the typical behaviour of a photon-counting detector, as discussed below. The corresponding "noise" is defined as the standard deviation of the reconstructed values $\mathbf{P}_M X(\mathbf{r})$ at a given point $\mathbf{r}$. The SNR is then

$$SNR[\mathbf{P}_M X](\mathbf{r}) = \frac{\overline{\mathbf{P}_M X}(\mathbf{r})}{[\mathrm{Var}(\mathbf{P}_M X)(\mathbf{r})]^{1/2}}, \qquad (13)$$

where $\mathrm{Var}(\mathbf{P}_M X)(\mathbf{r}) = \overline{\left[\mathbf{P}_M X(\mathbf{r}) - \overline{\mathbf{P}_M X}(\mathbf{r})\right]^2}$ is the corresponding (noise) variance. Using Eq.(4), we find that

$\overline{\left[\mathbf{P}_M X(\mathbf{r}) - \overline{\mathbf{P}_M X}(\mathbf{r})\right]^2} = \overline{[\sum_{m=1}^{M}(a_m - \overline{a}_m)U_m(\mathbf{r})]^2} = \sum_{m=1}^{M}\sum_{m'=1}^{M}\mathrm{Cov}(a_m, a_{m'})U_m(\mathbf{r})U_{m'}(\mathbf{r})$, where

$\mathrm{Cov}(a_m, a_{m'}) = \overline{a_m a_{m'}} - \overline{a}_m \overline{a}_{m'}$ are the covariances of the measured coefficients. We also assume that the measurements of different coefficients $a_m$ are statistically independent and hence $\mathrm{Cov}(a_m, a_{m'}) = \delta_{mm'}\mathrm{Var}(a_m)$, and therefore, $\mathrm{Var}(\mathbf{P}_M X)(\mathbf{r}) = \sum_{m=1}^{M}\mathrm{Var}(a_m)U_m^2(\mathbf{r})$. Therefore,

$$SNR[\mathbf{P}_M X](\mathbf{r}) = \sum_{m=1}^{M}\overline{a}_m U_m(\mathbf{r}) / \left[\sum_{m=1}^{M}\mathrm{Var}(a_m)U_m^2(\mathbf{r})\right]^{1/2}. \qquad (14)$$

In order to estimate the noise in the measured coefficients $a_m$, we use the conventional semi-classical model of statistical optics [26]. According to this approach, the propagation of light is calculated for continuous deterministic electromagnetic fields, and the quantization and randomness is considered in connection with light sources and photodection. For most thermal and similar light sources, to a very good approximation, the dominant contribution to noise in the registered signals comes from photodetection shot noise and, possibly, also from other detector-related noises (e.g. electronic dark current noise) [26]. For simplicity, we consider here only the case of a perfect photon-counting detector, for which only the photon shot noise is significant.

A photon-counting detector converts the incident radiant energy (integrated over the detector's entrance surface) into the corresponding photon numbers, with the efficiency of the process described by the previously introduced detection efficiency constant $\eta$ which has the dimensionality of photons per Joule [26]. In this process of conversion, a deterministic flux of light energy entering the detector results in the stochastic process of photon counting satisfying Poisson statistics [26]. Therefore, the mean values of the measured coefficients $a_m = \eta I_{in} \iint\limits_{\Omega} X(\mathbf{r}) T_m(\mathbf{r}) d\mathbf{r}$ are equal to $\overline{a}_m = \overline{n} \, x_m$, where $\overline{n} = \eta \, I_{in} |\Omega|$ is the mean number of photons used in the measurement of each coefficient, and $x_m = \dfrac{1}{|\Omega|} \iint\limits_{\Omega} X(\mathbf{r}) T_m(\mathbf{r}) d\mathbf{r}$ are dimensionless transmission coefficients. Similarly, according to the properties of Poisson statistics of the shot noise, $\mathrm{Var}(a_m) = \overline{a}_m = \overline{n} x_m$.

Let $\overline{N} \equiv M \overline{n}$ be the mean total number of photons utilized in the whole experiment (this already takes into account the detection efficiency). Then, in view of Eq.(4), Eq.(14) can be rewritten as a product of two distinct factors:

$$SNR[\mathbf{P}_M X](\mathbf{r}) = (\overline{N} / M)^{1/2} F_{M,X}(\mathbf{r}), \tag{15}$$

where the first factor, $(\overline{N} / M)^{1/2}$, reflects the effect of the photon statistics on the SNR, and the second term,

$$F_{M,X}(\mathbf{r}) \equiv \left( \sum_{m=1}^{M} x_m S_m(\mathbf{r}) \right) \Big/ \left( \sum_{m=1}^{M} x_m S_m^2(\mathbf{r}) \right)^{1/2} \tag{16}$$

is a deterministic function of $\mathbf{r}$, where, by definition, the vectors $S_m \equiv S_m(\mathbf{r})$ constitute a biorthogonal basis for $T_m(\mathbf{r})$: $< S_m, T_{m'} > = \delta_{mm'}$. We will call $F_{M,X}(\mathbf{r})$ the form factor. This factor depends only on the set of transmission masks $T_m(\mathbf{r})$ and the object transmission function $X(\mathbf{r})$, but not on the number of photons. Note that while the vectors $\boldsymbol{T}_m$ correspond to the original illumination vectors normalized by the mean number of incident photons, i.e. $\boldsymbol{T}_m = \bar{n}^{-1}\boldsymbol{W}_m$, the vectors $\boldsymbol{S}_m$ represent the suitably normalized versions of the biorthogonal vectors $\boldsymbol{U}_m$: $\boldsymbol{S}_m = \bar{n}\,\boldsymbol{U}_m$.

Consider the case, when the illumination vectors $W_m \equiv \eta \,|\,\Omega\,|\,I_m(\mathbf{r})$, $m = 1, 2, ..., M$, are orthogonal and have the same length with respect to the scalar product defined in Eq.(1), so that $<W_m, W_{m'}> = w^2 \delta_{mm'}$, where $w = \bar{n}t$ and $t = \|\,T_m\,\|$ for all $m$. Here we have $S_m = t^{-2}T_m$, $<S_m, T_{m'}> = t^{-2}\bar{n}^{-2} <W_m, W_{m'}> = \delta_{mm'}$, which leads to an expression for the form factor in terms of the orthogonal transmission masks:

$$F_{M,X}(\mathbf{r}) = \left(\sum_{m=1}^{M} x_m T_m(\mathbf{r})\right) \Big/ \left(\sum_{m=1}^{M} x_m T_m^2(\mathbf{r})\right)^{1/2}. \qquad (17)$$

The form factor in Eq.(17), and hence also the corresponding SNR, still generally depend on the object transmission function, $X(\mathbf{r})$, as well as on the transmission masks $T_m(\mathbf{r})$, and can have different values at different points $\mathbf{r}$ inside $\Omega$. In practice, when the performance of an imaging system is evaluated, the SNR is often measured in "flat" (uniform) areas of the reconstructed images, which are much larger than the area of the system's PSF. As the effect of uniform attenuation, $X(\mathbf{r}) = const$, in a "flat" object is trivial, we shall only consider the SNR in the case of $X(\mathbf{r}) = 1$, i.e. in the absence of the object. Considering the definition of the SNR in Eq.(13) with $X(\mathbf{r}) = 1$, we note first that, as a consequence of Condition 1, $\overline{\mathbf{P}_M 1}(\mathbf{r}) = 1$. Since for arbitrary illumination patterns the noise variance in the reconstruction of a uniform object may still be position-dependent, we shall spatially average the value of the noise variance, $\text{Var}(\mathbf{P}_M 1)(\mathbf{r})$, in the denominator of Eq.(13). This leads to the notion of squared averaged "flat" SNR, $SNR_a^2$, defined as the ratio of the spatially-averaged squared reconstructed signal to the spatially-averaged noise variance, in the absence of an object:

$$SNR_a^2 \equiv \frac{<(\overline{\mathbf{P}_M 1}(\mathbf{r}))^2>}{<\text{Var}(\mathbf{P}_M 1)(\mathbf{r})>} = (\bar{N}\,/\,M)(F_{M,1}^a)^2. \qquad (18)$$

In the derivation of Eq.(18), we used Eq.(15) and introduced the squared "flat" averaged form-factor,

$$(F^a_{M,1})^2 \equiv 1 / \left( \sum_{m=1}^{M} t_m \; \| \boldsymbol{S}_m \|^2 \right), \tag{19}$$

which is defined by spatially averaging the squared denominator of Eq.(16), with $X(\mathbf{r}) = 1$ and $t_m \equiv \; <\boldsymbol{T}_m>$. Note that the flat SNR defined in Eq.(18) does not depend on the imaged object, and hence is an intrinsic characteristic of the imaging system.

As vectors $\boldsymbol{S}_m$ are biorthogonal to $\boldsymbol{T}_m$, and, in particular, $<\boldsymbol{S}_m, \boldsymbol{T}_m> = 1$, the length of vectors $\boldsymbol{S}_m$ must be larger than or equal to the inverse of the length of $\boldsymbol{T}_m$, i.e. $\| \boldsymbol{S}_m \| \geq t^{-1}$. Therefore $\sum_{m=1}^{M} t_m \| \boldsymbol{S}_m \|^2 \geq t^{-2} \sum_{m=1}^{M} t_m$, equality being achieved only in the case when the length of all vectors $\boldsymbol{S}_m$ is equal to $t^{-1}$. The latter is possible only when all vectors $\boldsymbol{S}_m$ are parallel to $\boldsymbol{T}_m$, i.e. in the orthogonal case, and the corresponding result can also be obtained directly from Eq.(17). Therefore, $SNR^2_a \leq (\bar{N} / M) \tilde{t}_M$, where the quantity $\tilde{t}_M \equiv t^2 / \sum_{m=1}^{M} t_m$ is a particular form of average transmission coefficient for a given set of transmission masks. Note that $\tilde{t}_M = \sum_{m=1}^{M} t_m^2 / \sum_{m=1}^{M} t_m$ in the orthogonal case, because $1 = \sum_{m=1}^{M} t_m <\boldsymbol{S}_m> = t^{-2} \sum_{m=1}^{M} t_m^2$ as a consequence of Condition 1. When $\tilde{t}_M$ is multiplied by the mean number of incident photons, it makes the $SNR^2_a$ equal to the suitably averaged number of photons registered in each individual measurement, in the case of orthogonal illumination patterns and no object. This result agrees with the naturally expected behaviour in the case of image noise dominated by photon shot noise. Note that this SNR does not increase with the number of measurements, as the measurements of individual bucket coefficients are independent. Instead, when the reconstruction makes use of all $M$ individual measurements, the increased number of measured bucket coefficients is translated into a larger number of "effective pixels" in the reconstructed distribution, i.e. it results in improved spatial resolution, rather than in an increased SNR in each "pixel".

Thus, we have shown that squared reconstructed flat SNR, $SNR^2_a$, is always equal to or smaller than the average number of photons registered in each individual measurement of a bucket coefficient, with the equality achieved only in the case of orthogonal illumination patterns. The fact that, in the case of non-orthogonal illumination patterns, the SNR is smaller than the value expected in the case of uncorrelated Poisson statistics, is related to the presence of effective spatial correlations between the data obtained with individual illumination patterns, even though the measurements of individual bucket coefficients $a_m$ are statistically independent. This phenomenon is studied further in Sections 5-8 below.

It has been previously argued [19,20] that the ratio of SNR to spatial resolution, divided by the square root of the incident photon fluence, provides a good measure of the "quality" of an imaging system. The latter ratio, which was called the "intrinsic quality characteristic" (IQC) in [19], is also an invariant of the system which does not change under linear filtering [20]. Therefore, it is useful to estimate this characteristic in the case of computational imaging systems considered in the present paper. It follows from Eqs.(11) and (15) that, when the space $V_M(\Omega)$ contains constant functions (i.e. under Condition 1), we have

$$Q_2(\mathbf{r}) \equiv \left(\frac{|\Omega|}{\bar{N}}\right)^{1/2} \frac{SNR[\mathbf{P}_M X](\mathbf{r})}{(\Delta_2 r)[g_{\mathbf{r}}]} = F_{M,X}(\mathbf{r}) \left(\sum_{m=1}^{M} V_m^2(\mathbf{r}) / M\right)^{1/2}. \tag{20}$$

The subscript index "2" in the notation $Q_2(\mathbf{r})$ reflects the fact that the spatial resolution is calculated according to Eq.(10), rather than on the basis of the spatial variance of the PSF [20]. When the Green's function is shift-invariant (i.e. under Condition 2), we have $\sum_{m=1}^{M} V_m^2(\mathbf{r}) = M$ , and hence

$$Q_2(\mathbf{r}) = F_{M,X}(\mathbf{r}) . \tag{21}$$

As we noted above, the form factor $F_{M,X}(\mathbf{r})$ is independent from the number of photons used in the imaging experiment, but it can be spatially distributed differently for different illumination masks and imaged objects.

An object-independent "flat" version of the IQC can be defined in terms of the flat SNR, as per Eqs.(18)-(19):

$$Q_{2,a} = F_{M,1}^{a} \leq (\tilde{t}_M)^{1/2} , \tag{22}$$

with $F_{M,1}^{a}$ defined in Eq.(19) and $\tilde{t}_M = t^2 / \sum_{m=1}^{M} t_m$ . When the illumination vectors satisfy the orthogonality condition $<\mathbf{W}_m, \mathbf{W}_{m'}> = w^2 \delta_{mm'}$, it follows that $Q_{2,a} = (\tilde{t}_M)^{1/2}$.

When one is interested in estimating the efficiency of an imaging system with respect to the radiation dose delivered to the sample (which is proportional to the photon fluence incident on the object rather than on the masks), it can be useful to consider the quantity

$$\tilde{Q}_{2,a} \equiv (\bar{N} / \bar{N}_t)^{1/2} Q_{2,a} \leq M^{1/2} t / \sum_{m=1}^{M} t_m , \qquad (23)$$

where $\bar{N}_t \equiv \bar{N} M^{-1} \sum_{m=1}^{M} t_m$ is the mean total number of photons incident on the object (after transmission through the masks) during the measurements of $M$ bucket coefficients.

It is also possible to calculate the squared flat SNR in the measured signal $Y_{1,M}(\mathbf{r})$ corresponding to $X(\mathbf{r}) = 1$, by evaluating

$SNR_{a,in}^2 \equiv < \overline{Y_{1,M}(\mathbf{r})}>^2 / < \mathrm{Var}(Y_{1,M}\mathbf{r}) > = (\bar{N} / M) \sum_m \sum_{m'} t_m t_{m'} < \mathbf{T}_m, \mathbf{T}_{m'} > / \sum_m t_m \| \mathbf{T}_m \|^2$. This

value becomes equal to $(\bar{N} / M) \tilde{t}_M$ in the orthogonal case, i.e. when $< \mathbf{T}_m, \mathbf{T}_{m'} > = t^2 \delta_{mm'}$. This means, in particular, that in the case of orthonormal illumination patterns the flat SNR does not change in the reconstruction process. As we showed previously that the spatial resolution also does not change between the measurement and the object spaces in this case, it follows that the flat IQC $Q_{2,a} = (\tilde{t}_M)^{1/2}$ is invariant under the action of the reconstruction operator $\mathbf{R}_M$ in the case of orthogonal illumination patterns.

## 5. Localized illumination masks

Let us consider an example which is based on a conventional imaging setup with a pixelated detector with $L \times L$ pixels (for which the issues of spatial resolution, SNR and IQC have been previously considered using similar criteria [20]), but presented here in a form consistent with the imaging setup with structured illumination and single-pixel detector as considered in Section 2 of this paper. For this purpose, we define a set of $M = L^2$ single-pixel illumination patterns, $V_m(\mathbf{r}) = LT_m(\mathbf{r})$, where $L$ is an integer and $T_m(\mathbf{r}) = T(\mathbf{r} - \mathbf{r}_m)$ are single-pixel transmission masks. The points $\mathbf{r}_m = (m_x h, m_y h)$ denote the positions of the centres of the detector pixels with the area $h^2$ each, indexed by integers $m = (m_x - 1)L + m_y$,

$m_x, m_y = 1, 2, ..., L$, $m = 1, 2, ..., M$. The function $T(\mathbf{r})$ is the "indicator function" of a pixel, i.e. $T(x, y) = 1$, when $|x| < h/2$ and $|y| < h/2$, and $T(x, y) = 0$ is zero otherwise. Each shifted function $V_m(\mathbf{r}) = LT(\mathbf{r} - \mathbf{r}_m)$ is then equal to $L$ inside the pixel with index $m$, represented by the domain $\Omega_m = \{ |x - m_x h| < h/2, |y - m_y h| < h/2 \}$, and is equal to zero elsewhere. The whole image domain is represented by the square

$\Omega = \bigcup_{m=1}^{M} \Omega_m = \{ h/2 < x < Lh + h/2, h/2 < y < Lh + h/2 \}$, consisting of $M = L^2$ pixels $\Omega_m$,

with the total area $|\Omega| = M |\Omega_m| = L^2 h^2$. The orthonormality relationship, $< V_m, V_{m'} > = \delta_{mm'}$

for any $m,m' = 1,2,...,M$, with the scalar product defined by Eq.(1), follows from the fact that the spatial supports of functions $V_m(\mathbf{r})$ with different indexes do not overlap, while for $m = m'$ we get: $|\Omega|^{-1} \iint\limits_{\Omega} V_m^2(\mathbf{r}) d\mathbf{r} = |\Omega|^{-1} L^2 |\Omega_m| = 1$.

The projector onto the space spanned by vectors $V_m$, $m=1,2,...,M$, can be defined as in Eq.(5). The corresponding Green's function is not shift-invariant in general (although it is invariant with respect to shifts by a whole number of pixels). The spatial resolution can be evaluated in accordance with Eq.(10):

$$(\Delta_2 r)^2 [g_{\mathbf{r}}] = |\Omega| / \sum_{m=1}^{M} V_m^2(\mathbf{r}) = |\Omega| / L^2 = |\Omega| / M = h^2 , \qquad (24)$$

where we have taken into account the fact that any point $\mathbf{r}$ in $\Omega$ lies within a single pixel, and hence there is one and only one basis function which is equal to $L$ at this point (here we ignore the points $\mathbf{r}$ that lie on the boundaries of pixels, because such points constitute a set of measure zero). We see that the spatial resolution here is the same at any point $\mathbf{r}$. The result agrees with one's natural expectation that the spatial resolution of the "direct" imaging setup using a detector with the pixel size $h$, should be equal to that pixel size.

The SNR can be calculated according to Eqs.(15)-(16). Because in this case the original illumination patterns, $W_m = \bar{n} T_m = (\bar{n} / L) V_m$ are orthogonal, the form factor $F_{M,X}(\mathbf{r})$ can be expressed by Eq.(17) with $x_m = \dfrac{1}{|\Omega|} \iint\limits_{\Omega} X(\mathbf{r}) T(\mathbf{r} - \mathbf{r}_m) d\mathbf{r}$. Note also that because the supports of indicator functions of different pixels do not overlap, we have $T_m(\mathbf{r}) T_{m'}(\mathbf{r}) = \delta_{mm'} T_m^2(\mathbf{r})$ for any point $\mathbf{r}$ in $\Omega$. Using this property, it is easy to see that $\left[ \sum_{m=1}^{M} x_m T_m(\mathbf{r}) \right]^2 = \sum_{m=1}^{M} x_m^2 T_m^2(\mathbf{r}) = x_{m(\mathbf{r})}^2$ and $\sum_{m=1}^{M} x_m T_m^2(\mathbf{r}) = x_{m(\mathbf{r})}$, where $m(\mathbf{r})$ is the index of the pixel containing the point $\mathbf{r}$ and $x_{m(\mathbf{r})} = \dfrac{1}{|\Omega|} \iint\limits_{\Omega} X(\mathbf{r}') T_{m(\mathbf{r})}(\mathbf{r}') d\mathbf{r}'$. Substituting these expressions into Eq.(17), we obtain that $F_{M,X}^2(\mathbf{r}) = x_{m(\mathbf{r})}$, and therefore

$$SNR^2[\mathbf{P}_M X](\mathbf{r}) = (\bar{N} / M) x_{m(\mathbf{r})} , \qquad (25)$$

which corresponds to the average number of (transmitted) photons registered in the pixel containing the point $\mathbf{r}$. Note that $x_{m(\mathbf{r})} = M^{-1} < X(\mathbf{r}) >_m$, where $< X(\mathbf{r}) >_m = \frac{1}{|\Omega_m|} \iint\limits_{\Omega} X(\mathbf{r}')T_{m(\mathbf{r})}(\mathbf{r}')d\mathbf{r}'$ is the average value of the object transmission function

over the pixel containing the point $\mathbf{r}$. The additional factor $M^{-1}$ here reflects the fact that incident photons are used very inefficiently when single-aperture masks and a bucket detector are used instead of uniform illumination (no masks) and a position-sensitive detector. A similar result is obtained for the flat SNR: $SNR_a^2 = (\bar{N}/M)(t^2/\sum_{m=1}^{M} t_m) = \bar{N}/M^2$, since

$t^2 = t_m = \frac{1}{|\Omega|} \iint\limits_{\Omega_m} dr = 1/M$ in this case. Note, however, that as far as the radiation dose

delivered to the sample is concerned, the present method is as efficient as the direct imaging with a pixelated detector. Indeed, the average transmission coefficient of the masks is equal to $M^{-1} \sum_{m=1}^{M} t_m = M^{-1}$ here. Therefore, the SNR calculated with respect to the number of photons transmitted through the masks (i.e. the photons incident on the object, $\bar{N}_t = \bar{N}/M$ ) is equal to $SNR_a^2 = \bar{N}_t/M$, which is the same as in the case of imaging with a pixelated detector, no masks and the total number of incident photons equal to $\bar{N}_t$. Moreover, if the scattering in the object is significant, the transmission data obtained with the help of a single-pixel transmission mask and a bucket detector can represent a more accurate estimate of the projected transmission of the object along a given one-pixel-wide "pencil" ray, compared to the data obtained with a plane-wave illumination and a pixelated detector. Indeed, in the latter case the photons scattered during the transmission of one such ray through the object can contribute to the signal measured in several adjacent pixels of the detector. This phenomenon represents a well-known advantage of the "first generation" computed tomography [14].

Equations (24) and (25) lead to the following result for the IQC:

$$Q_2^2(\mathbf{r}) = \frac{|\Omega|}{\bar{N}} \frac{\bar{N} \, x_{m(\mathbf{r})}}{M \, h^2} = x_{m(\mathbf{r})} = \frac{< X(\mathbf{r}) >_m}{M}. \tag{26}$$

If we recalculate the latter quantity with respect to the photons incident on the object, i.e. divide $Q_2^2(\mathbf{r})$ by the average transmission coefficient of the masks, $M^{-1} \sum_{m=1}^{M} t_m = M^{-1}$, we will arrive at the result $\tilde{Q}_2^2(\mathbf{r}) \equiv MQ_2^2(\mathbf{r}) = < X(\mathbf{r}) >_m$. In other words, the IQC of the system with single-pixel masks, evaluated with respect to the incident dose delivered to the sample, is equal to the square root of the absorption coefficient of the object. Considering the case $X(\mathbf{r}) = 1$ we obtain that $Q_{2,a}^2 = M^{-1}$ and $\tilde{Q}_{2,a} = 1$. This agrees with the result previously derived in [20] by a different approach.

Consider now the imaging system with overlapping two-pixel transmission masks $T_m^{(2)}(\mathbf{r}) = T_m(\mathbf{r}) + T_{m+1}(\mathbf{r})$, $m = 1,2,...,M$, where $M$ is odd and the index $m$ is formally considered cyclical, i.e. $T_{M+1}(\mathbf{r}) \equiv T_1(\mathbf{r})$. Such overlapping masks are no longer orthogonal, as $< T_m^{(2)}, T_{m+1}^{(2)} > = 1/M$. It is straightforward to verify that the biorthogonal basis here consists of the vectors $S_m^{(2)}(\mathbf{r}) = (M/2)[\sum_{m'=1}^{m} (-1)^{m-m'} T_{m'}(\mathbf{r}) + \sum_{m'=m+1}^{M} (-1)^{m+1-m'} T_{m'}(\mathbf{r})]$. Therefore, $\| S_m^{(2)} \|^2 = (M^2/4) \sum_{m'=1}^{M} \| T_{m'} \|^2 = M^2/4$. Evaluating the squared flat form factor according to Eq.(19) and substituting it into Eq.(22), we obtain that the squared flat IQC is equal to $Q_{2,a}^2 = 1 / \left( \sum_{m=1}^{M} t_m^{(2)} \| S_m^{(2)} \|^2 \right) = 1 / \left( \sum_{m=1}^{M} (2/M)(M^2/4) \right) = 2/M^2$. The corresponding IQC corrected for the average mask transmission coefficient $M^{-1} \sum_{m=1}^{M} t_m^{(2)} = 2/M$ is equal to $\tilde{Q}_{2,a}^2 = 1/M$. We see that both these squared IQCs are smaller, by factors proportional to $1/M$, than the corresponding characteristics for the single-mask system considered above. This is an instructive example of the effect of non-orthogonality of illumination patterns on the SNR and IQC of imaging systems. Note that the object-space spatial resolution of the system with overlapping two-pixel masks $T_m^{(2)}(\mathbf{r})$ is the same as the one for the system with single-pixel masks $T_m(\mathbf{r})$, as expected in the case of "super-resolution" imaging with sub-pixel shifting of a PSF. However, as we see from the above results, the SNR achieved in such super-resolution imaging will be considerably smaller compared to the SNR in the single-pixel imaging system, at the same incident photon fluence. In other words, in order to achieve the same image quality (i.e. the same SNR and the same spatial resolution) as in the system with the PSF of area $h^2$, by sub-pixel shifting of the PSF of area $2h^2$, one would have to increase the radiation dose delivered to the imaged object by a factor of $M$, where $Mh^2$ is the field of view.

## 6. Delocalized harmonic masks

Here we consider a computational imaging system defined as in Section 2, but with the harmonic transmission masks $T_m(\mathbf{r}) = 1/2 + F_m(\mathbf{r})$, $F_m(\mathbf{r}) = f_{m_x}(x) f_{m_y}(y)$, $f_1(t) = 1/2$, $f_l(t) = (1/\sqrt{2}) \sin(\pi l t / A)$ when $l = 2,4,...$, is even, and $f_l(t) = (1/\sqrt{2}) \cos(\pi(l-1)t/A)$, when $l = 3,5,...$, is odd. Here $m = (m_x - 1)L + m_y$, $m_x, m_y = 1,2,...,L$, and $m = 1,2,...,M$. The image domain is represented by the square $\Omega = \{-A/2 < x < A/2, -A/2 < y < A/2\}$. Note that we normalized all Fourier harmonics in the above definition of transmission masks in a way that ensures that the transmission values are non-negative and do not exceed 1 at any point $\mathbf{r}$ in $\Omega$. The illumination vectors here are $W_m = \bar{n} T_m$, where $\bar{n} = \eta I_{in} |\Omega|$ is the average number of photons used in the measurement of each bucket coefficient, as defined above. The orthonormalized basis can be chosen as follows: $V_1 = (4/3)\overline{n}^{-1}W_1 = (4/3)T_1 = 1$ and $V_m = 4\overline{n}^{-1}[W_m - (2/3)W_1] = 4[T_m - (2/3)T_1] = 4F_m$. It is straightforward to verify that $<V_m, V_{m'}> = \delta_{mm'}$, $m, m' = 1, 2, \ldots, M$.

Note that $\sum_{m=1}^{M} V_m^2(\mathbf{r}) = M$, provided that $M = 1 + 4M'$ for some positive integer $M'$, which is easy to show by grouping together terms with the same $m_x$ and $m_y$. Restricting the values of $M$ to integers of the form $1 + 4M'$, we obtain that the spatial resolution evaluated in accordance with Eq.(10) will be equal to

$$(\Delta_2 r)^2[g_\mathbf{r}] = |\Omega| / \sum_{m=1}^{M} V_m^2(\mathbf{r}) = |\Omega| / M \ . \tag{27}$$

This shows, as one would naturally expect, that the spatial resolution here is equal to the image area divided by the number of Fourier harmonics used in the imaging system.

Using the simple relation $<F_m, F_{m'}> = (1/16)\delta_{mm'}$, it is straightforward to verify by direct calculations that in this case the biorthogonal basis $U_m$, such that $<U_m, W_{m'}> = <U_m, \overline{n}(1/2 + F_{m'})> = \delta_{mm'}$, consists of functions $U_1(\mathbf{r}) = \overline{n}^{-1}(4/3)[1 - 8\sum_{m=2}^{M} F_m(\mathbf{r})]$, $U_m(\mathbf{r}) = \overline{n}^{-1}16F_m(\mathbf{r})$, $m = 2, 3, \ldots, M$. Taking into account that $S_m(\mathbf{r}) = \overline{n} U_m(\mathbf{r})$ and substituting the obtained expressions for $S_m(\mathbf{r})$ into Eqs.(16) and (21), we obtain:

$$Q_2(\mathbf{r}) = F_{M,X}(\mathbf{r}) = \frac{x_1 + 4\sum_{m=2}^{M}(3x_m - 2x_1)F_m(\mathbf{r})}{\left[x_1[1 - 16\sum_{m=2}^{M}F_m(\mathbf{r})] + 16\sum_{m=2}^{M}\sum_{m'=2}^{M}(9x_m\delta_{mm'} + 4x_1)F_m(\mathbf{r})F_{m'}(\mathbf{r})\right]^{1/2}} \ . \tag{28}$$

Using Eq.(28) with $x_m = t_m$ and taking into account that $<F_m> = 0$, when $m > 1$, we obtain the following value for the flat IQC:

$$Q_{2,a}^2 = (F_{M,1}^a)^2 = t_1^2 / \left[t_1 + \sum_{m=2}^{M}(9t_m + 4t_1)\right] = 3 / \left[4 + 40(M-1)\right] . \tag{29}$$

Dividing $Q_{2,a}^2$ by the average transmission coefficient of the masks,

$$M^{-1}\sum_{m=1}^{M}t_m = [(3/4 + (1/2)(M-1))]/M \cong 1/2 \text{, when } M >> 1 \text{, we obtain that}$$

$\tilde{Q}_{2,a}^2 \cong 3/[2 + 20(M-1)]$ for large $M$.

One can see that here the value of the IQC, calculated with respect to the incident dose delivered to the sample, rapidly decreases when the number of illumination patterns $M$ grows. This can be contrasted with the constant value $\tilde{Q}_{2,a} = 1$ obtained for general orthogonal illumination patterns at the end of Section 5. This difference is a typical consequence of non-orthogonality of illumination patterns. The low value of the IQC here is related to the fact that the average signal in the considered computational imaging setup receives contribution from only one measurement, i.e. that of coefficient $a_1$, while the noise variances from measurements of all coefficients $a_m$, $m=1,...,M$, add up. In other words, the measurements of all non-zero coefficients $a_m$ do not contribute to the average signal, but they do contribute to the noise. Note, however, that the measurements of non-zero coefficients $a_m$ are still not "useless", as far as the quality measure $Q_{2,a}$ is concerned, as these coefficients contribute to the spatial resolution, as is evident in Eqs.(11)-(12).

### 7. Pseudo-random delocalized masks

This example has direct relationship to standard ghost imaging using random speckle fields [25], computational ghost imaging [9-12] and compressive sensing [21].

In this example we assume that the first illumination pattern is spatially uniform, $I_1(\mathbf{r}) \equiv I_1 = const$, and the other illumination patterns satisfy the equation [25]

$$\frac{1}{|\Omega|}\iint_{\Omega}[I_m(\mathbf{r}) - I_1][I_{m'}(\mathbf{r}) - I_1]d\mathbf{r} = \sigma^2\delta_{mm'}, \ \ m,m' = 2,3,...,M, \tag{30}$$

where the spatial average of the intensity patterns is assumed to be the same,

$I_1 = \frac{1}{|\Omega|}\iint_{\Omega}I_m(\mathbf{r})d\mathbf{r}$, for all $m=1,...,M$, and $\sigma^2 = |\Omega|^{-1}\iint_{\Omega}[I_m(\mathbf{r}) - I_1]^2 d\mathbf{r}$ is the spatial variance

of the intensity patterns, assumed to be the same for all $m > 1$ (the spatial variance of the pattern $I_1(\mathbf{r}) = const$ is obviously equal to zero). In view of Eq.(30), the orthonormalized vectors $\mathbf{V}_m$ can be naturally chosen as

$$V_1(\mathbf{r}) = \frac{I_1(\mathbf{r})}{I_1} = 1 , \ V_m(\mathbf{r}) = \frac{I_m(\mathbf{r}) - I_1}{\sigma} , \ m = 2,3,\ldots,M. \tag{31}$$

If the set of illumination patterns is shift-invariant, i.e. it satisfies Condition 2, then the spatial resolution of this imaging system is given by Eq.(12). In order to estimate the SNR using Eqs.(15)-(17), we need to find the biorthogonal basis $\boldsymbol{U}_m$ or the coefficients $q_{ml}$. Introducing the vectors $\boldsymbol{F}_m \equiv \boldsymbol{I}_m / I_1 - 1 = \boldsymbol{T}_m / t_1 - 1 , \ m = 2,3,\ldots,M , \ t_1 = <T_1> = T_1$, we obtain from Eq.(30) that $<\boldsymbol{F}_m, \boldsymbol{F}_{m'}> = \kappa^{-2}\delta_{mm'}$, where $\kappa = I_1 / \sigma \geq 1$. It is then straightforward to verify by direct calculations that in this case the biorthogonal basis $\boldsymbol{U}_m$ , such that

$<\boldsymbol{U}_m, \boldsymbol{W}_{m'}> = <\boldsymbol{U}_m, \overline{n} t_1 (1 + \boldsymbol{F}_{m'})> = \delta_{mm'}$, consists of functions $U_1(\mathbf{r}) = \overline{n}^{-1} t_1^{-1} [1 - \kappa^2 \sum_{m=2}^{M} F_m(\mathbf{r})]$,

$U_m(\mathbf{r}) = \overline{n}^{-1} t_1^{-1} \kappa^2 F_m(\mathbf{r}) , \ m = 2,3,\ldots,M$ . Substituting $S_m(\mathbf{r}) = \overline{n} U_m(\mathbf{r})$ into Eqs.(16) and (21), we obtain:

$$Q_2(\mathbf{r}) = F_{M,X}(\mathbf{r}) = \frac{x_1 + \kappa^2 \sum_{m=2}^{M} (x_m - x_1) F_m(\mathbf{r})}{\left\{ x_1[1 - 2\kappa^2 \sum_{m=2}^{M} F_m(\mathbf{r})] + \kappa^4 \sum_{m=2}^{M} \sum_{m'=2}^{M} (x_m \delta_{mm'} + x_1) F_m(\mathbf{r}) F_{m'}(\mathbf{r}) \right\}^{1/2}} . \tag{32}$$

As in the case of harmonic illumination masks, the form factor here can in general be different for different transmission functions and different points in the image domain. The squared flat IQC in this example is equal to

$$Q_{2,a}^2 = (F_{M,X}^a)^2 = t_1^2 / \left[ t_1 + \kappa^2 \sum_{m=2}^{M} (t_m + t_1) \right] = t_1 / \left[ 1 + 2\kappa^2(M-1) \right], \tag{33}$$

where we have taken into account that the average transmission of all masks is assumed to be the same, i.e. $t_m = t_1$ for any $m$. Dividing $Q_{2,a}^2$ by the average transmission coefficient of the masks, which is equal to $t_1$, we obtain that $\tilde{Q}_{2,a}^2 \cong 1 / \left[ 1 + 2\kappa^2(M-1) \right]$. Therefore, the value of the IQC, calculated with respect to the incident dose delivered to the sample, rapidly decreases when the number of illumination patterns $M$ grows. As in the case with overlapping two-pixel masks and harmonic delocalized masks considered above, this behaviour is a consequence of non-orthogonality of illumination patterns.

Because the IQC of imaging systems with pseudo-random delocalized masks rapidly deteriorates as the number of illumination patterns increases, it appears that it can only remain reasonably high if the number of measurements $M$ can be kept low. This effectively requires that the class of imaged objects is strongly correlated with a small number of illumination patterns, ensuring that the representation of any unknown imaged object $X(\mathbf{r})$ is "sparse" in the basis of vectors $I_m(\mathbf{r})$. As this is exactly the central premise of compressive sensing imaging [15,21], the method can be "saved" by this sparsity assumption. Indeed, the present argument can be viewed as motivating the necessity to use compressive sensing approaches in computational ghost imaging of scenes with $M \gg 1$ resolution elements using $M \gg 1$ pseudo-random illumination patterns.

## 8. Discussion and conclusions

The major difference in the estimated behaviour of the SNR in the case of computational imaging using orthogonal illumination patterns, as in the case of single-pixel masks, and the non-orthogonal delocalized harmonic masks or the pseudo-random illumination patterns satisfying Eq.(30), can be attributed to the non-orthogonality in the latter cases. The SNR significantly deteriorates when the coefficients $c_m$ used for the reconstruction of the unknown object transmission function according to Eq.(4), are calculated from the experimentally measured coefficients $a_m$ containing random noise. A decrease in SNR inevitably takes place when one subtracts two measured values, such as $a_m$, $m > 1$, and $a_1$, because the mean values of the coefficients are subtracted, reducing the magnitude of the resultant "signal", but the variances add up, increasing the resultant noise. This unfavourable effect is well known in many forms of imaging that involve post-detection processing of experimentally registered light intensity distributions. For example, it is exactly this effect that is known to lead to low SNR and correspondingly large low-frequency artefacts in in-line phase-contrast imaging with the Transport of Intensity equation [28-30], when the signal of interest is obtained by subtraction of two images collected at different defocus distances.

The considerations presented in the previous paragraph can in fact be applied to a generalized model of imaging systems described by Eq.(4), where the signal of interest is reconstructed by means of an equation which is linear with respect to the measured values, $a_m$, which contain Poisson or other uncorrelated random noise. Note that the measured coefficients $a_m$ can correspond to intensities measured at individual pixels of a detector or to integrated intensities collected with a structured illumination pattern and a single-pixel (bucket) detector, or to something else still. In fact, we can drop the reference to the method by which the input values $a_m$ are measured, and also ignore the particular methods of image reconstruction, leaving just a general linear equation connecting the measured intensity values $a_m$ and the "reconstructed object" coefficients $c_m$ in the same vector basis:

$$c_m = \sum_{m'=1}^{M} r_{mm'} a_{m'} \,, \tag{34}$$

where $r_{mm'} \equiv q_{mm'}^{(2)}$ according to Eq.(4). The corresponding continuous version of Eq.(34) is given by a linear integral equation with the Green's function $R_M(\mathbf{r},\mathbf{r}') = \sum_{m=1}^{M} U_m(\mathbf{r})U_m(\mathbf{r}')$.

Depending on the qualitative way the SNR and the spatial resolution are affected by Eq.(34), one can naturally divide such systems into three distinct classes:

(I) "Rotation and multiplication like" class. In this case the matrix $r_{mm'}$ is a product of an orthogonal matrix and a constant number, and the corresponding integral transform is proportional to a unitary operator. The simplest example can be given in the case $M = 2$ by the transformation $c_1 = a_1 + a_2$, $c_2 = a_2 - a_1$, which can be represented as a rotation by the angle equal to $\pi/4$ radians, followed by multiplication by the factor $\sqrt{2}$. In this case, the image reconstruction operation does not change the average SNR achieved in the measured signal (a decrease of SNR for some coefficients $c_m$, compared to the coefficients $a_m$, is exactly matched by the corresponding increase for the other coefficients) and also does not change the spatial resolution (it preserves the width of the spatial Fourier spectrum).

(II) "Convolution-like" class. In this case the matrix elements $r_{mm'}$ are typically all non-negative. The corresponding integral transform acts in a way similar to low-pass filtering. The simplest example can be given in the case $M = 2$ by the transformation $c_1 = a_1$, $c_2 = (a_1 + a_2)/2$. In this case, the image reconstruction operation increases the average SNR (reduces noise by means of spatial correlations), but worsens spatial resolution (shrinks the spatial Fourier spectrum), in the transition from the measurement space (corresponding to coefficients $a_m$) to the object space (corresponding to coefficients $c_m$).

(III) "Deconvolution-like" class. In this case the matrix elements $r_{mm'}$ typically have alternating signs, and the corresponding integral transform acts somewhat similar to high-pass filtering. The simplest example can be given in the case $M = 2$ by the transformation $c_1 = a_1$, $c_2 = 2a_2 - a_1$, which is inverse to the transformation given as an example for Class II above. In the present case, the image reconstruction operation reduces the average SNR (increases noise by means of spatial "decorrelations"), but improves the spatial resolution (broadens the spatial Fourier spectrum).

The imaging system with single-pixel masks considered in Section 5 belongs to Class I. In such systems, the spatial resolution and the average SNR do not change in the transition from the measurement space to the object space. Examples of systems from Class II are represented by any low-pass filtering operation. Such systems improve the SNR (by means of spatial correlations), but lower the spatial resolution (blur the image). The imaging systems with overlapping two-pixel masks considered at the end of Section 5, the harmonic masks considered in Section 6, and the pseudo-random masks considered in Section 7, all belong to Class III. The average SNR in this case decreases in the transition from the measured bucket coefficients $a_m$ to the object coefficients $c_m$. The spatial resolution does improve in the transition from the image (measurement) space to the object space. The latter fact was recently demonstrated in an experiment [25].

Interestingly, the situation with the imaging systems using non-orthogonal illumination patterns, as considered above, is exactly opposite to the situation with reconstructive imaging using the homogeneous Transport of Intensity equation (TIE-Hom) [30], as considered in our recent publication [31]. It has been shown previously that the TIE-Hom imaging is capable of increasing SNR by factors of up to two orders of magnitude, without sacrificing spatial resolution, in certain common imaging contexts [32-36]. We showed in [31] that the IQC can increase in the process of free-space propagation of a transmitted wave from the object space to the measurement space (i.e. the image / detector space), because here the spatial resolution can improve without an increase of the noise. The key to that fact is the behaviour of photon noise, which, in the case of thermal light sources, is dominated by the photodetection shot noise. Given that the total number of photons is preserved in the process of free-space propagation, the average photon shot noise is the same in the object and the image spaces (possibly, after the geometric magnification is taken into account). Thus the noise stays the same, while the spatial resolution can improve (the spatial Fourier spectrum can become broader) in the process of free-space propagation in near-Fresnel region [31], giving one a net increase in the ratio of SNR to the spatial resolution ratio. In the case of imaging with non-orthogonal illumination patterns, the situation is the opposite. As we have demonstrated above, in that case the ratio of the SNR to the spatial resolution (and the corresponding IQC) in the object space is lower in the case of non-orthogonal illumination patterns compared to the equivalent orthogonal case. This happens because the spatial resolution in the object space is the same for orthogonal and non-orthogonal (linearly independent) bases, as the spatial resolution in the reconstructed object is determined only by the dimensionality of the space spanned by the illumination vectors, as shown in Section 3. At the same time, the SNR is lower in the case of non-orthogonal bases, compared to orthogonal bases with the same average transmission, as shown in Section 4, because in the non-orthogonal cases the SNR is drastically lowered (typically in proportion to the number of illumination patterns) by the reconstruction operator which effectively "spatially decorrelates" the noise in the measured signal (because Eq.(34) performs a deconvolution in this case).

Note that unlike the situation with the imaging setups considered in [20], in the setup studied in the present paper it was not possible to show that the IQC is the same in the measurement and object spaces, i.e. that it does not change in the process of object reconstruction. This is probably due to the fact that although the reconstruction operator $\mathbf{R}_M$ defined in Eq.(8) is linear, it is not necessarily shift-invariant when the illumination patterns are not orthogonal. The behaviour of the spatial resolution and SNR in the case of measurements in non-orthogonal bases, and its consequences for the efficiency of the corresponding imaging systems, may be worth investigating further in the future.

## Acknowledgements

The authors are grateful to David Ceddia for useful discussions.